\documentclass{elsart}

\usepackage{natbib}
\usepackage{graphics}
\usepackage{amssymb}

\begin{document}
\def\EE#1{\times 10^{#1}}
\def\Msun{{~\rm M}_\odot}
\def\kms{\rm ~km~s^{-1}}
\def\lsim{\!\!\!\phantom{\le}\smash{\buildrel{}\over
  {\lower2.5dd\hbox{$\buildrel{\lower2dd\hbox{$\displaystyle<$}}\over
                               \sim$}}}\,\,}
\def\gsim{\!\!\!\phantom{\ge}\smash{\buildrel{}\over
  {\lower2.5dd\hbox{$\buildrel{\lower2dd\hbox{$\displaystyle>$}}\over
                               \sim$}}}\,\,}

\def\EE#1{\times 10^{#1}}
\def\kms{\rm ~km~s^{-1}}
\def\msunyr{~M_\odot~{\rm yr}^{-1}}
\def\Mdot{\dot M}

\begin{frontmatter}

\title{Optical and Infrared Observations of Radioactive Elements 
in Supernovae.}

\author{Jesper~Sollerman}

\address{European Southern Observatory, Karl-Schwarzschild-Strasse 2,
D-85748 Garching bei M\"unchen, Germany}

\begin{abstract}
At late phases the powering of supernova light curves is often
provided by the decay of radioactive elements synthesized in the
explosions. 
This is unambiguously revealed when the light curve decline follows
the half life time of the decaying elements, and the bolometric
luminosity then directly provides the mass of ejected radioactive
material. I will focus on the best observed element, $^{56}$Ni, and
demonstrate that different supernovae eject different amounts of this element.
SN 1994W ejected very small amounts of nickel, 
possibly caused by black hole formation. SN 1998bw
may instead have ejected more $^{56}$Ni than any other
supernova to date.
I will also discuss
our ISO non-detection of [Fe II] $26~\mu$m in SN 1987A, which can be used
to estimate an upper
limit on the mass of ejected radioactive $^{44}$Ti.
\end{abstract}

\end{frontmatter}

\section{Introduction}
Optical and infrared
observations of core-collapse supernovae provide crucial clues about
the formation of radioactive nuclei.
Although we read in astronomy text books that the elements in the
universe are constantly created and expelled in supernova explosions, 
the hard, observational, evidence for this story may not be as compelling as
we would like (e.g., Kirshner 1996).
Determining the ejected masses of radioactive material from the energy
budget in late light curves is one of the most direct pieces of
evidence 
that explosive nucleosynthesis is indeed taking place in supernovae.
I will 
present two supernovae
that ejected substantially different amounts of $^{56}$Ni than did the
'canonical' SN 1987A. 
I will also report on our non-detections of SN 1987A with the
Infrared Space Observatory, and how these observations
were used to obtain an upper limit on the amount of ejected
radioactive $^{44}$Ti.

\section{Nickel masses in core-collapse supernovae}

If the optical depth in the ejecta is high enough to trap the
energetic gamma-rays 
from the decaying radioactive elements, the late supernova light curve
decline will reflect the half life time of the element
decay rate. The most important radioactive isotopes for supernova
powering are $^{56}$Ni, $^{57}$Ni, and $^{44}$Ti, 
dominating at successively later phases.
Only for SN 1987A have all three
decays been observed to power the light curve.
Observations of other SNe usually cover
only the $^{56}$Ni phase, as revealed in the bolometric light curve 
by a fading of 0.98 magnitudes every 100 days.
With a handle on the gamma-deposition, we need to determine the bolometric
luminosity to deduce the mass of ejected nickel. 
This has
been determined for a dozen supernovae.
Table 1 shows an incomplete sample, 
obtained with varying degrees of sophistication. 
Patat et al. (1994) presented 8 Type IIP SNe with late time
absolute V-magnitudes. Here, I have assumed 
the same bolometric correction as for SN 1987A, 
i.e., a linear scaling with $0.071\Msun$ of $^{56}$Ni.
This gives a
mean value for these 8 SNe of $0.075\Msun$ of $^{56}$Ni 
with a standard deviation of only $0.03\Msun$.
In fact, the number 0.07 occurs rather frequently in Table 1. 
Though there exist no theoretical
justification for this number, only the Australian group around Brian Schmidt
seemed to find
supernovae with varying yields of nickel.

\section{The low amount of $^{56}$Ni in SN 1994W}

But not all supernovae eject the same amount of $^{56}$Ni. 
SN 1994W
ejected significantly less nickel than the canonical
0.07$\Msun$ (Sollerman, Cumming $\&$ Lundqvist 1998,
hereafter SCL98). This
discovery was later followed by another low-nickel supernova, SN
1997D (Turatto et al. 1998).
This is important for understanding the galactic
iron-enrichment, but may also tell us something about the
explosion mechanism itself, and possibly even about the nature of the
compact remnant formed in the supernova explosion.

We know that fall-back is an important process taking place after
a successful shock has been launched in the supernova explosion. 
The innermost material
falling back onto the newly formed neutron star will
largely be made up of $^{56}$Ni. 
It is also clear that a substantial amount of fall-back may trigger
a collapse of the neutron star into a black hole.
The quantitative features of this scenario may be
difficult to calculate, but the main consequences are clear.
A supernova for which fall-back is efficient enough to form a black
hole will have a diminished amount of radioactive material ejected.
An underluminous light curve tail from a Type IIP supernova may therefore be
the signature of a supernova that formed a black hole 
(Woosley $\&$ Timmes 1996).
SN 1994W, discovered on 1994 July 29 in NGC 4041 may indeed be such a case.
At the plateau stage this supernova was very luminous, but then showed an
unprecedented drop in the light curve (see Fig. 1 in SCL98). 
In 12 days the supernova faded
by 3.5 magnitudes, and then continued to fade rapidly on the light
curve tail.

The absolute R-band magnitudes of SN 1994W are significantly
fainter than those of the well observed SN 1987A.
This tells us that SN 1994W ejected less
radioactive nickel than the canonical $0.07\Msun$.
The exact amount of nickel somewhat depends on the interpretation of the
fast decline rate on the light curve tail, as detailed in SCL98. 
However, regardless of the interpretation, 
the mass of ejected $^{56}$Ni was very low, $<~0.015\Msun$. In fact, the
observations are consistent with a scenario where all nickel
fell back onto the neutron star, perhaps causing it to implode to a 
black hole,
and where the late emission is due to 
explosion energy diffusing out of the denser core/mantle region
(Woosley $\&$ Weaver 1986).

Clearly this is not an unambiguous proof of black hole formation. However,
if a massive star would form a black hole due to
extensive fall back, the observational consequence should be a light
curve that looks pretty much like that observed for SN
1994W.
These discoveries
have triggered some interesting theoretical investigations.
Zampieri et al. (1998) have discussed the
prospects of directly observing the black hole accretion luminosity 
emerging at
late phases for supernovae with low input of radioactive heating.
Keeping an observational eye on also the faint supernovae may thus be 
rewarding.

\section{SN 1998bw - The most nickel rich supernova to date}
SN 1998bw was probably associated
with the Gamma-ray burst GRB 980425.
Early modeling implied a very energetic explosion ejecting some
$0.5-0.7\Msun$ of $^{56}$Ni, an order of magnitude more than for SN 1987A
(Iwamoto et al. 1998; Woosley, Eastman \& Schmidt 1999). However,
H\"oflich et al. (1999) noted that if 
SN 1998bw were an asymmetric explosion the
early light curve may give an erroneous estimate of the nickel-mass. 
We obtained late time observations
to model the supernova at phases when the ejecta are optically thin,
and the effects of asymmetry should be minimal (Sollerman et al. 2000).

Using the massive CO-star models of Iwamoto et al. (1998) and Woosley et
al. (1999), we modeled the late time emission in detail. 
To successfully reproduce the peaked line profiles in the spectra we had to
macroscopically mix the original structure and 
increase the inner densities. Having a handle on the distribution of
material in the supernova, we could then accurately calculate the
gamma-deposition and thus determine the nickel-mass required
to power the late light curve.
We obtained
$0.5\Msun$ in one model, and $0.9\Msun$ in the
other. Independent of the exact number, 
simple arguments show that SN 1998bw ejected 
at least $0.3\Msun$ of $^{56}$Ni, 
significantly more than the $0.07\Msun$ of SN 1987A.
Together with SN 1994W, it is thus clear that core-collapse supernovae
can eject very different amounts of nickel.

Due to the high velocities the ejecta of SN 1998bw became
almost completely transparent to the gamma-rays at later phases, and
the energy input should eventually be dominated by the kinetic energy from the
positrons emitted in the radioactive decays. Then the nickel-mass is
directly given by the bolometric luminosity. A preliminary bolometric
light curve is reported in Patat et al. (2001) and modeled by Nakamura
et al. (2001). We are currently trying to improve on this.

This is because the background of SN 1998bw is very complex and even
PSF-subtraction is likely to overestimate the luminosity of the SN at
late phases. Figure 1 shows three subsequent 
HST/STIS images of the region with a scale
of merely $1.5''\times1.5''$. It is clear that several objects will
fall into any ground based PSF.
These images also show that one of the objects is still fading,
confirming the SN identification made by Fynbo et al. (2000).
To assess the problem of background subtraction in greater detail
we have obtained late
VLT template images to subtract from our earlier images, a technique
utilized in the High-z supernova context (Fig. 2). 

Moreover, including our late VLT-IR photometry 
and the HST photometry we can construct a better
late time light curve to address the importance of positron deposition at
this very late stage. A preliminary attempt is shown in figure 3.

%

\section{The amount of $^{44}$Ti in SN 1987A}

The emission of
SN 1987A at very late times ($\gtrsim1500$ days) is powered by the decay of
$^{44}$Ti. This nucleus is also 
produced in the very center of the supernova, and thus
probes the early phases of the explosion.

At these very late phases most of the supernova light 
emerges outside the bands observable from ground. Only a few
percent of the emission emerges in the UVOIR and constructing
a bolometric light curve from such observations is utterly difficult. 
The most sensitive way to determine the powering at this
stage should instead be in the far-IR. The emission line of [Fe
II] $26~\mu$m appears to be the ideal probe and 
accounts for almost half of the supernova emission at these phases.
This line was observable by the Infrared Space Observatory (ISO) which
was therefore pointed toward SN 1987A. However, no
emission from the supernova was detected (Lundqvist, Sollerman, Kozma
et al. 1999). From the upper limit on this emission, together with
detailed modeling of the
supernova, we achieved an upper limit on the
amount of titanium powering the late emission. Including a number of
possible uncertainties (atomic data, distance, mixing etc.)
resulted in an upper limit of $^{44}$Ti of 
$<1.5\times~10^{-4}\Msun$ from the non-detection at day 3999.

However, 
Borkowski et al. (1997) indicated
that another set of ISO data reached a ten times lower limit, severely 
challenging estimates from explosion models.
We analyzed these data when they 
became available from the ISO data archive.
Again, no emission from the
supernova was observed, but the obtained upper limit on the flux was 
indeed lower.
With an updated
code for the modeling, we arrived at an upper limit of
$\sim 8\times~10^{-5}\Msun$ from the non-detection at day 3425
(Lundqvist, Kozma, Sollerman $\&$ Fransson 2001).

In the new investigation we furthermore analyzed several systematic
effects that could affect the far-IR lines. We found
that a certain type of positron-escape from the iron-rich gas to the
macroscopically mixed Si-gas rich regions
could decrease the flux in the $26~\mu$m line by some $30\%$ without
significantly altering the optical bands that are indeed well
reproduced by the model.
Including this effect we conclude that no more than
$\sim 1.1\times~10^{-4}\Msun$ of $^{44}$Ti was ejected from SN 1987A.


\clearpage

\begin{table}
\caption{Some nickel yields from core-collapse supernovae}
\label{nitable}

\[
         \begin{tabular}{lcl}
            \hline
            \noalign{\smallskip}
Supernova & M($^{56}$Ni) & Source \\
\noalign{\smallskip}
            \hline
            \noalign{\smallskip}

SN 1969L & 0.083~$\Msun$ & Patat et al. (1994) \\
SN 1979C & 0.060 & Patat et al. (1994) \\
SN 1980K & 0.036 & Patat et al. (1994) \\
SN 1983K & 0.10 & Patat et al. (1994) \\
SN 1985P & 0.039 & Patat et al. (1994) \\
SN 1986I & 0.13 & Patat et al. (1994) \\
SN 1988A & 0.069 & Patat et al. (1994) \\
SN 1987A & 0.071 & Suntzeff \& Bouchet (1990) \\
SN 1990E & 0.073& Schmidt et al. (1993) \\
SN 1992am & 0.30 & Schmidt et al. (1994) \\
SN 1991G & 0.024 & Blanton et al. (1995) \\
SN 1994I & 0.07 & Young et al. (1995)  \\
SN 1992H & $\sim0.075$ & Clocchiatti et al. (1996) \\
SN 1993J & 0.08 & Houck \& Fransson (1996) \\

\noalign{\smallskip}
            \hline
         \end{tabular}
      \]

\end{table}

\clearpage

\begin{figure*}
\includegraphics{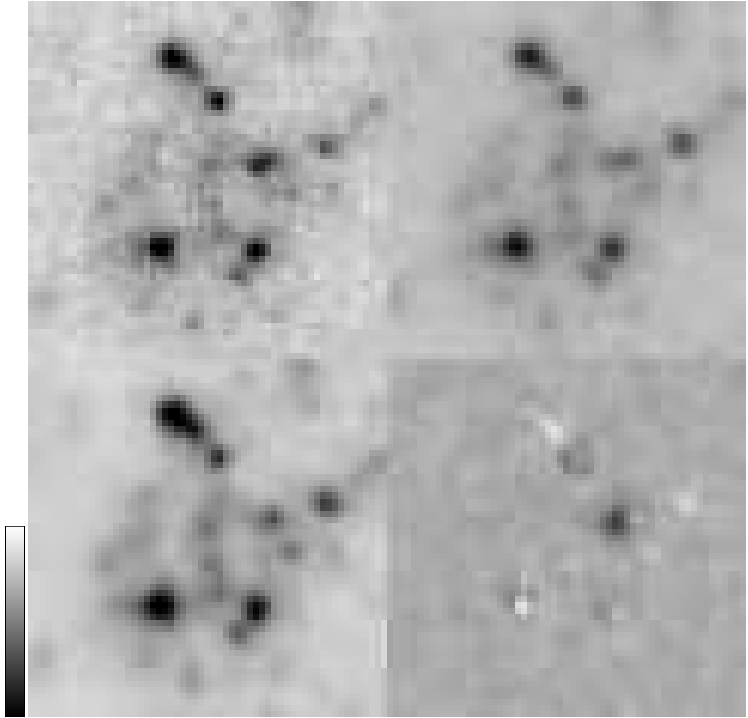}
\vspace{5.5cm}
\caption{
HST/STIS CLEAR images of SN 1998bw and its environment. 
The images are obtained
in June 2000 (upper left), November 2000 (upper right), 
and in August 2001 (lower left). The
scale is merely $1.5''\times1.5''$. Lower right shows a subtraction between
the first and the final frame. The supernova is clearly identified.
(Courtesy of the Supernova INtensive Study-team featuring S. Holland.)
}
\label{sins.ps}
\end{figure*}

\clearpage

\begin{figure*}
\includegraphics{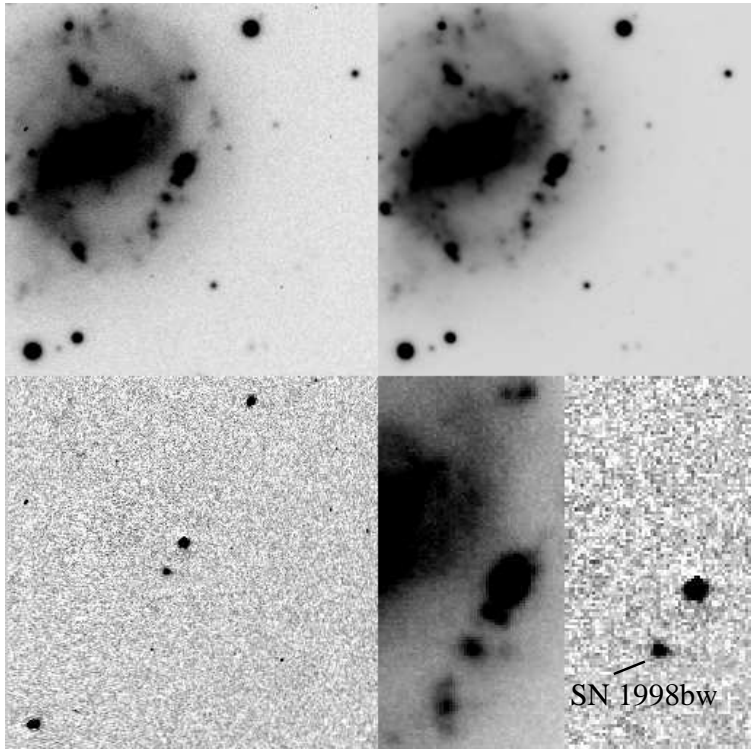}
\vspace{9. cm}
\caption{
The template subtraction method. Upper left shows the
supernova and its galaxy in September 1999, 503 days past
explosion. Upper right is the template image. The subtraction is shown
in the lower left. Note that only the supernova is seen, as well as
some residuals from saturated stars.
Lower right shows a blowup of the supernova region, before and
after subtraction. 
}
\label{fig98bw.ps}
\end{figure*}

\begin{figure*}
\includegraphics{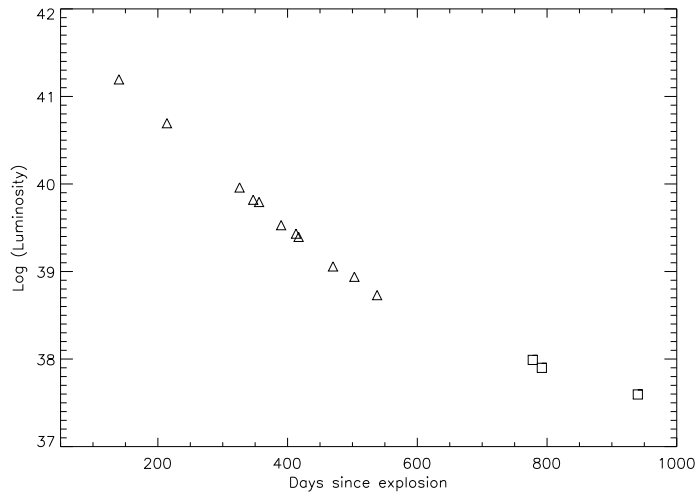}
\vspace{5.5 cm}
\caption{
Preliminary bolometric light curve from the template subtraction method. Late
time observations from VLT (triangles, L$_{\rm BVRI}$) and HST (squares, 
L$_{\rm CL}$).
{\bf Added:} The final version of this light curve 
is published in Sollerman et al. 2002 
(A\&A, 386, 944), where a possible interpretation of the very late light curve in terms of powering of more long-lived isotopes is given.}
\label{bolom.ps}
\end{figure*}

\end{document}